\documentclass[aps, prb, reprint, superscriptaddress, floatfix]{revtex4-2}
\usepackage{graphicx}
\usepackage{dcolumn}
\usepackage{bm}
\usepackage{physics}
\usepackage{amsfonts}
\usepackage{mathrsfs}
\usepackage{subfigure}
\usepackage{hyperref}
\usepackage{color}
\usepackage{xcolor}
\DeclareMathAlphabet{\pazocal}{OMS}{zplm}{m}{n}
\usepackage{amsmath}
\usepackage{amssymb}
\usepackage[T1]{fontenc}

\begin{document}

\preprint{APS/123-QED}

\title{Photo-induced pattern formations and melting of charge density wave order}

\author {Lingyu Yang}
\affiliation{Department of Physics, University of Virginia, Charlottesville, Virginia, 22904, USA}

\author {Gia-Wei Chern}
\affiliation{Department of Physics, University of Virginia, Charlottesville, Virginia, 22904, USA}

\begin{abstract}
We investigate the out-of-equilibrium dynamics of a photo-excited charge-density-wave (CDW) state in the square-lattice Holstein model, in a setup similar to a pump-probe experiment. At half-filling, the ground state of this system is characterized by a checkerboard modulation of particle densities, accompanied by a concomitant lattice distortion. An efficient real-space dynamics method integrating the von~Neumann equation for electron density matrix and Newton equation for classical lattice dynamics is developed to simulate the dynamical evolution of a photo-excited Holstein system. We find that the energy injected by a short pump pulse results in the reduction of the CDW order and the generation of coherent phonons. At strong photoexcitations, while the CDW order is melted in the sense that the time-averaged order parameter vanishes, a dynamical CDW state is self-sustained by the strong coherent phonon oscillations. Our large-scale simulations further uncover a dynamical regime of intermediate fluence where complex pattern formation is induced by the pump pulse. The emergent spatial textures are characterized by super density modulations on top of the short-range checkerboard CDW order. Our demonstration of pattern formation highlights the significance of dynamical inhomogeneity in quantum many-body systems in pump-probe experiments.
\end{abstract}


\date{\today}

\maketitle

\section{Introduction}

\label{sec:intro}

The nonequilibrium time evolution of quantum systems has garnered significant attention in recent years. This surge in interest is fueled in part by remarkable experimental progress, particularly in ultrafast techniques such as pump-probe spectroscopy~\cite{Othonos98,Shah99,Tsen04,ulbricht11,Fischer16,Zong2023}. By exciting a sample with a short laser pulse (pump pulse), this technique allows one to study the ultrafast relaxation dynamics of quasiparticles~\cite{Rossi02,Kira11}. Recently, pump-probe techniques have also been utilized to investigate photo-induced ultrafast collective behaviors~\cite{Matsunaga_2012, Matsunaga_2013,Mansart_2013,Matsunaga_2014, Polli_2007, Beaud_2014, Hashimoto_2014}, for example, by measuring the time-resolved dynamics of order-parameter fields. Moreover, photo-induced phase transition~\cite{Nasu_2004,Tokura06} with an intensive pump pulse offers an avenue to detect long-lasting meta-stable states or even the so-called hidden states which are nonequilibrium many-body states without an equilibrium counterpart~\cite{Kiryukhin_1997,Chollet_2005,Onda_2008,  Fausti_2011, Ichikawa_2011, Stojchevska_2014, Zhang_2016, Mitrano_2016,Giannetti_2016, Nova_2019, Li_2019}.

Charge density wave (CDW) states are one of the prominent platforms for exploring ultrafast phenomena associated with a symmetry breaking phase. This is partly due to the ubiquity of CDW states which have been observed in a wide class of materials~\cite{Gruner_1988, Thorne_1996}. Since many CDW states are stabilized by electron-phonon coupling, CDW order is often found to compete or coexist with a proximate superconducting state.  Moreover, as the charge degrees of freedom directly couples to external electric field, CDW materials are idea candidates for manipulating and tracing photo-excited phase transitions. For example, photo-induced insulator-to-metal transitions through the melting of CDW order have been observed in many materials~\cite{Iwai_2007, Kubler_2007, Schmitt_2008, Tomeljak_2009, Hellmann_2010, Kawakami_2010,Cavalleri_2004, Schmitt_2008, Tomeljak_2009, Hellmann_2010, Petersen_2011, Wall_2012, Liu_2013, Mathias_2016, Rettig_2016, Chavez-Cervantes_2019, Diego_2021,Yusupov_2010, Porer_2014,Rohwer_2011,Hellmann_2012,Mertelj_2013}.

While a CDW order can be stabilized through a purely electronic mechanism, the majority of CDW states are accompanied by a concomitant structural distortion. This points to the important role of electron-phonon coupling in both the static and dynamical behaviors of CDW states. In particular, the dynamics of CDW order is also intimately related to that of lattice degrees of freedom. The interplay between the photo-excited electron-hole pairs and the combined CDW and structural order could lead to rich dynamical phenomena. For example, it has been shown that the melting of a CDW is often accompanied by the generation of coherent phonons~\cite{Rohwer_2011, Hellmann_2012, Porer_2014, Hedayat_2019, Sayers_2020}. 

Phenomenologically, the CDW dynamics is usually modeled by the time-dependent Ginzburg-Landau theory with either real or complex order parameter fields~\cite{McMillan75,Mihailovic_2013,Zong19,Trigo19,Dolgirev20}. The effective energy functional of a CDW order parameter can be viewed as obtained by integrating out the electron degrees of freedom. While such empirical approach might capture the general adiabatic CDW dynamics, it cannot properly account for the ultrafast processes that involve photoexcited electron-hole pairs and their interplay with the collective CDW behaviors. On the other hand, ultrafast CDW dynamics have also been simulated using many-body techniques ranging from time-dependent Hartree-Fock (TDHF)~\cite{Seo18,yang24} and nonequilibrium Green's functions~\cite{Schuler18} to computationally very demanding nonequilibrium dynamical mean-field theory (DMFT)~\cite{Shen14,Matveev16,Freericks17,Murakami15,Petocchi23} and time-dependent generalizations of the density matrix renormalization group (DMRG) methods~\cite{Hashimoto17,Stolpp_2020}.

Due to the computational complexity, most of these numerical studies are restricted to well-defined model systems. For example, the nonequilibrium DMFT was used to obtain exact photoemission of the melting of CDW order in the Falicov-Kimball model in the infinite dimension limit~\cite{Shen14,Matveev16,Freericks17}. The nonequilibrium nature of the photoexcited CDW order manifests itself in the emergence of a gapless transient state yet with a finite charge modulation. The time-evolving block decimation (TEBD) algorithm based on matrix-product states was employed to simulate the scenario of CDW melting in one-dimensional (1D) Holstein model~\cite{Hashimoto17,Stolpp_2020}, which is another canonical system that hosts CDW states. These works uncovered the importance of domain-walls or solitons, which are unique topological defects for 1D systems, in the destruction of the CDW order. 


The soliton-driven scenario of CDW melting in 1D also highlights the crucial role of topological defects and spatial inhomogeneity in photo-induced nonequilibrium states. Indeed, due to the locality principle, relaxation of a perturbed system often proceeds in a spatially incoherent manner. This naturally leads to inhomogeneous intermediate states, similar in spirit to the Kibble-Zurek mechanism~\cite{Kibble_1976, Zurek_1985}. However, the effects of photo-induced spatial fluctuations and heterogeneous structures in the melting of higher-dimensional CDW order remain unexplored. This is partly due to the intrinsic difficulty of multi-scale dynamical modeling of many-body quantum systems. For example, reliable TEBD or other time-dependent DMRG simulations are restricted to 1D systems, while spatial fluctuations are ignored in the nonequilibrium DMFT approaches. 

Another intriguing possibility is the emergence of complex patterns in order-parameter fields induced by a laser pulse. For timescales shorter than the quasiparticle relaxation time, the coherent dynamics of order parameter fields is governed by a nonlinear system of coupled von~Neumann equations for electrons and Newton dynamics for lattice. A hallmark of highly nonlinear dynamical systems is the formation of complex patterns. Indeed, dynamically self-organized structures have been extensively studied for decades in contexts such as reaction-diffusion systems and Turing instability~\cite{Nicolis77,Cross_1993,Koch_1994,Pismen06,Cross09}. However, much less is known about similar pattern formation phenomena in out-of-equilibrium quantum systems. For example, it remains to be seen whether some of the unifying themes and universal behaviors of pattern formation in classical systems can be applied to nonequilibrium quantum systems.

In this paper, we present large-scale real-space simulations of photo-induced ultrafast CDW dynamics in a 2D semi-classical Holstein model~\cite{Holstein59}. This model is a prototypical system for studying phenomena related to electron-phonon coupling, such as phonon-mediated superconductivity~\cite{Costa18,Bradley21}, polaron dynamics~\cite{Bonca99,Mishchenko14}, and in particular CDW physics~\cite{Noack91,Zhang19,Chen19,Esterlis19}. Also notably, CDW orders in the Holstein model are intimately related to lattice distortion, thus providing a platform for investigating the interplay between collective CDW behaviors and lattice dynamics. To account for dynamical inhomogeneities induced by laser pulses, an efficient real-space methods is developed by combining the von~Neumann equation for the electron density matrix with the Newton equation for the lattice degrees of freedom. 

We find that while the CDW order is reduced due to the injected energy, the  photoexcitation also generates a coherent oscillation of both the CDW order and lattice distortion. The melting of the CDW order depends critically on the amplitude as well as the frequency of the laser pulse. Our large-scale simulations show that the melting process proceeds through the breakup of the initially homogeneous CDW domain into a highly inhomogeneous state with a finite CDW order locally. Intriguingly, we find that for intermediate laser fluences the system exhibits stripe modulations of the CDW amplitude, which in some cases evolve to more complex patterns at later times. At strong photoexcitations, we further observe that spatial inhomogeneity is suppressed by a large coherent phonon oscillation, giving rise to a self-sustained dynamical CDW order.

The rest of the paper is organized as follows. In Sec.~\ref{sec:model}, we briefly review the Holstein model and the Peierls substitution for modeling the laser pulse excitation. We also discuss the governing equations of the CDW state in the semi-classical Holstein model and the efficient implementation of the real-space method. The ultrafast dynamics of photo-induced CDW states is summarized in Sec.~\ref{sec:melting_order}. A systematic analysis of the CDW order on the fluence and frequency of the pump pulse is also presented. Detailed descriptions of the pattern formation and its structures in some specific frequency are summarized in Sec.~\ref{sec:pattern}.  Finally, we conclude the paper with a summary and outlook in Sec.~\ref{sec:conclusion}.

\section{Model and Methods}

\label{sec:model}

We consider a Holstein model~\cite{Holstein59} with spinless fermions on a square lattice, described by the following Hamiltonian with three parts
\begin{equation}
	\hat{\mathcal{H}}=\hat{\mathcal{H}}_{\mathrm{e}} + \hat{\mathcal{H}}_{\mathrm{L}} + \hat{\mathcal{H}}_{\mathrm{eL}}.
\end{equation}
The first term $\hat{\mathcal{H}}_{\mathrm{e}}$ describes the hopping of electrons between nearest neighboring sites
\begin{equation}
	\label{eq:H_e}
	\hat{\mathcal{H}}_{\mathrm{e}}= - t_{\rm nn} \sum_{\langle ij\rangle} \left(\hat{c}^{\dag}_{i} \hat{c}^{\,}_{j} + \hat{c}^{\dag}_{j} \hat{c}^{\,}_{i} \right),
\end{equation}
where $\hat{c}^\dagger_i$ ($\hat{c}^{\,}_i$) denotes the creation (annihilation) operator of an electron at lattice site-$i$. The second part describes a dispersion-less Einstein phonon model:
\begin{equation}
	\hat{\mathcal{H}}_{\mathrm{L}}=\sum_{i}\left( \frac{1}{2m} \hat{P}^{2}_{i}+\frac{1}{2}m\Omega^2 \hat{Q}^{2}_{i} \right)
\end{equation}
where $m$ is the effective mass, $\Omega$ is the intrinsic oscillation frequency, and $K\equiv m\Omega^2$ is the force constant of phonons. The third term describes a local coupling between the electron density $\hat{n}_i = \hat{c}^\dagger_i \hat{c}^{\,}_i$ and the phonon mode
\begin{equation}
	\label{eq:H_eL}
	\hat{\mathcal{H}}_{\mathrm{eL}}=-g\sum_{i}\left( \hat{n}_i -\frac{1}{2}\right) \hat{Q}_{i}
\end{equation}
Physically, the phonon $Q_i$ can be viewed as the breathing mode of an oxygen octahedron surrounding each lattice site-$i$ in a perovskite structure. The above coupling can be thought of as a local version of the deformation potential electron-phonon interaction. 

The Holstein model on various bipartite lattices exhibits a robust CDW order that breaks the sublattice symmetry~\cite{Noack91,Zhang19,Chen19,Esterlis19}. In the case of square lattice, the CDW order is characterized by a $\mathbf K = (\pi, \pi)$ checkerboard charge modulation. Unlike the superconducting phase, which requires a full quantum treatment, the CDW order remains robust even in the semiclassical approximation. Indeed, the semiclassical phase diagram of the CDW order obtained by a hybrid Monte Carlo method agrees very well with that obtained from determinant quantum Monte Carlo simulations~\cite{Esterlis19}. 
Here the semiclassical approximation is invoked to describe the dynamical evolution of photo-induced CDW states in the 2D Holstein model. A similar semiclassical dynamics method was recently employed to study the photo-emission and long-time behaviors of CDW states in the 1D Holstein model~\cite{Petrovic22}. Our focus here, on the other hand, is on dynamical phenomena related to spatial inhomogeneity and pattern formation.  

For an initial state of a homogeneous CDW order whose wave function is described by a Slater determinant, the dynamics of the semiclassical Holstein model can be exactly solved, at least numerically. To this end, we assume a product form for the quantum state of the system: $\ket{\Gamma(t)}=\ket{\Phi(t)} \otimes \ket{\Psi(t)}$, where $\ket{\Phi(t)}$ and $\ket{\Psi(t)}$ denote the phonon and electron wave-functions, respectively. The semi-classical approximation for the lattice subsystem amounts to a direct product wave function $\ket{\Phi(t)}=\prod_{i} \ket{\phi_{i}(t)}$ for the phonons. As a result, the expectation value of phonon operators, e.g. $\langle \Gamma(t) | \hat{Q}_i | \Gamma(t) \rangle$ reduces to $Q_i(t) \equiv \langle \phi_i(t) | \hat{Q}_i | \phi_i(t) \rangle$, and similarly for the momentum operators, $P_i(t) \equiv \langle \Gamma(t) | \hat{P}_i | \Gamma(t)\rangle = \langle \phi_i(t) | \hat{P}_i |\phi_i(t) \rangle$. 

To derive the equation of motion for the `classical' variables $Q_i(t)$ and $P_i(t)$, we consider the expectation of Heisenberg equation of motion, $d \langle \hat{P}_{i} \rangle/dt = -i \langle [ \hat{P}_{i}, \hat{\mathcal{H}} ] \rangle / \hbar$, where $\langle ... \rangle$ is the expectation value computed using the full wave-function $\ket{\Gamma(t)}$. Direct calculation of the commutators yields the coupled Hamiltonian dynamics
\begin{eqnarray}
	\label{eq:newton_eq}
	\frac{dQ_i}{dt} = \frac{P_i}{m}, \qquad \frac{dP_i}{dt} = g n_i - K Q_i,
\end{eqnarray}
which is equivalent to the Newton equation of motion for a simple Harmonic oscillator with an additional force $g n_i$ due to the coupling to electron density. Here the time-dependent electron density is given by $n_i(t) = \langle \Gamma(t) | \hat{n}_i | \Gamma(t) \rangle$. Thanks to the product form of the system quantum state, this can be reduced to $n_i(t) = \langle \Psi(t) | \hat{n}_i | \Psi(t)\rangle$, i.e. its value only depends on the many-electron wave function $|\Psi(t) \rangle$. It is worth noting that, while the electron and phonon quantum states are not entangled, the lattice dynamics is still coupled to the evolution of the electrons. 

The time evolution of the electron wave function is governed by Schr\"odinger equation $i \hbar \partial |\Psi\rangle / \partial t = \hat{\mathcal{H}} |\Psi \rangle$. Since the Holstein Hamiltonian is bilinear in fermion operators, the many-electron state remains in a Slater determinant throughout the dynamical evolution. Instead of directly evolving the Slater determinant, a more efficient approach is based on the dynamical equation for the single-particle density matrix, 
\begin{eqnarray}
	\label{eq:rho1}
	\rho_{ij}(t) = \bra{\Psi(t)} \hat{c}^\dagger_j \hat{c}^{\,}_{i} \ket{\Psi(t)}.
\end{eqnarray}
The on-site electron number, which is a driving force of the lattice dynamics in Eq.~(\ref{eq:newton_eq}), is readily given by the diagonal elements: $n_i(t) = \rho_{ii}(t)$. The dynamical equation for the density matrix can be derived from the Heisenberg equations for the creation and annihilation operators. A more intuitive approach is to consider the effective single-particle Hamiltonian $H_{ij}$ defined as
\begin{eqnarray}
\label{eq:first_quantized_H}
    \hat{\mathcal{H}}_{\mathrm{e}}(t) + \hat{\mathcal{H}}_{\mathrm{eL}}(t)
    = \sum_{ij} \hat{c}^\dagger_{i} \, H_{ij}[\{Q_i(t) \}]\, \hat{c}^{\,}_j,
\end{eqnarray}
The matrix elements of $H$ can be read from Eqs.~(\ref{eq:H_e}) and (\ref{eq:H_eL}):
\begin{eqnarray}
	H_{ij} = -t_{ij} - g\delta_{ij} Q_i(t)
\end{eqnarray}
where $t_{ij} = t_{\rm nn}$ for nearest-neighbor pairs $\langle ij \rangle$ and zero otherwise. Given the single-particle Hamiltonian, the evolution of the density matrix $\rho$ is governed by the von~Neumann equation $d\rho/ dt = (i/\hbar) [\rho, H]$. Explicitly, we have
\begin{eqnarray}
\label{eq:von-neumann}
	\label{eq:von-neumann_eq}
    i\hbar \frac{d\rho_{ij}}{dt} =  \sum_{k}\left( \rho_{ik}t_{kj} - t_{ik}\rho_{kj} \right) + g\left( Q_{j} - Q_{i} \right)\rho_{ij}. \quad
\end{eqnarray}

In order to model the laser excitation, the Peierls substitution is employed to incorporate the coupling to the electric field of a laser pulse. Consider a uniform linearly polarized electric field $\mathbf E(t) = -\partial \mathbf A / \partial t$, where $\mathbf A(t) = \hat{\mathbf e} A(t)$ is the time-varying vector potential, and $\hat{\mathbf e}$ is the polarization vector. In the presence of an electric field, the electron hopping integral acquires a phase factor
\begin{eqnarray}
\label{eq:peierls}
    t_{ij}c^{\dag}_{i}c_{j} \rightarrow  t_{ij}e^{i \mathbf A(t) \cdot (\mathbf r_i - \mathbf r_j)}c^{\dag}_{i}c_{j}.
\end{eqnarray}
In our simulations below, we assume a Gaussian function for the laser pulse
\begin{eqnarray}
\label{eq:vector_potential}
    A(t)= \mathcal{A} \exp\left[(t-t_0)^2/\sigma^2 \right]\cos{\left[\omega (t-t_{0})\right]},
\end{eqnarray}
where $\mathcal{A} $ represents the amplitude of the pulse, $t_0$ is the peak time, $\sigma$ is the pulse width, and $\omega$ is the center frequency of the pulse. By setting the lattice constant $a = 1$, the Peierls phase in Eq.~(\ref{eq:peierls}) is $\exp[i {A}(t) ]$ for the nearest-neighbor bonds. The amplitude parameter $\mathcal{A}$ is then dimensionless and can be viewed as the maximum phase angle caused by the pump pulse.

\begin{figure*}[t]
\includegraphics[width=1.99\columnwidth]{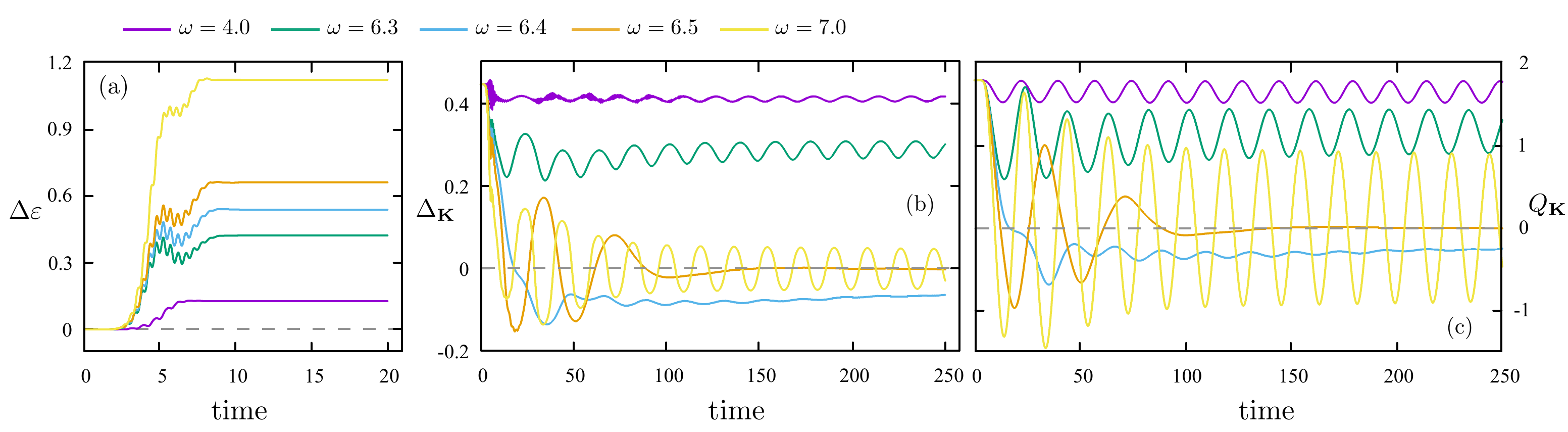}
\caption{\label{fig:n_omega} Time evoulution of (a) the injected energy per site~$\Delta \varepsilon(t) = [ E(t) - E_0 ]/N$, (b) the CDW order parameter $\Delta_{\mathbf K}(t)$, and (c) amplitude of staggered distortion $Q_{\mathbf K}(t)$, for laser pulse excitation with different center frequencies. The width of the laser pulse is fixed at $\sigma = 2$,  the peak time is at $t_0 = 5$, and the amplitude is $\mathcal{A} = 0.5$.}
\end{figure*}

There are two characteristic time scales for the dynamics of the Holstein model. First, from the bandwidth of the electron tight-binding model $W = 8 t_{\rm nn}$, one can define a time scale $\tau_{\rm e} = \hbar/ t_{\rm nn}$ for the electron dynamics. On the other hand, the natural frequency $\Omega$ of the Einstein phonons gives a characteristic time $\tau_{\rm L} = 1/ \Omega$ for the lattice dynamics.  The dimensionless adiabatic parameter is defined as the ratio $r = \tau_{\rm e} / \tau_{\rm L} = \hbar \Omega / t_{\rm nn}$. 
which characterizes the relative time scales of the two subsystems. A characteristic $Q^*$ for lattice distortions can be estimated from the balance of elastic energy and electron-phonon coupling: $K Q^{* 2} \sim g \langle n \rangle Q^*$. Assuming electron number $\langle n \rangle \sim 1$, we obtain $Q^* \sim g / K$. This in turn gives a momentum scale $P^* = m \Omega Q^*$ via Eq.~(\ref{eq:newton_eq}). Based on this characteristic scale, the electron-phonon coupling can be characterized by a dimensionless parameter $\lambda = g Q^* / W = g^2 / W K$. For all simulations discussed below, these two dimensionless parameters are set to $r = 0.4$ and $\lambda=1$. The simulation time is measured in unit of $\tau_{\rm e}$, energies are measured in units of $t_{\rm nn}$, and the lattice distortion is expressed in terms of $Q^*$.

\section{Photo-induced Melting of CDW order}

\label{sec:melting_order}

The method discussed in Sec.~\ref{sec:model}, namely the real-space von~Neumann equation~(\ref{eq:von-neumann_eq}) coupled with Newton equation~(\ref{eq:newton_eq}), is applied to simulate the photo-induced ultrafast CDW dynamics in a pump-probe setup. In all simulations below, we consider a polarization along the symmetric diagonal direction $\hat{\mathbf e} = (\hat{\mathbf x} + \hat{\mathbf y})/\sqrt{2}$, and a system size of $60 \times 60$. The system is initially prepared in a ground state with a homogeneous CDW order and a concomitant checkerboard lattice distortion. This initial CDW state is then subject to a short laser pulse of the wave form~(\ref{eq:vector_potential}). 
It is worth noting that the square-lattice tight-binding model exhibits a divergent Lindhard susceptibility at half-filling due to a perfect nesting of the Fermi surface. As a result, the system is unstable against the formation of a staggered lattice distortion $Q_i = \mathcal{Q} \exp(i \mathbf K \cdot \mathbf r_i)$, where $\mathbf K = (\pi, \pi)$ is the ordering wave vector of the checkerboard pattern and $\mathcal{Q}$ is the distortion amplitude. As detailed in Appendix~\ref{app:bogoliubov}, the staggered distortion induces a concomitant charge modulation and opens a spectral gap $\varepsilon_{\rm gap} = 2 g \mathcal{Q}$, rendering the system a CDW insulator. The energy gap is determined by the competition between the gain of electronic energy through gap opening and the cost of elastic energy.

The finite spectral gap also means that electron-hole pairs cannot be excited by photon energies less than the CDW band gap. This implies a threshold frequency $\hbar \omega_{\rm th} = \epsilon_{\rm gap} =  2g \mathcal{Q}$ for continuous wave excitations. Yet, instead of a sharp threshold transition as a function of frequency, a crossover behavior is expected due to a combination of nonlinear effects and finite width of the pump pulse. A laser pulse of width $\sigma$ comprises photons of energies in a finite range $\hbar(\omega \pm \delta\omega)$ around the center frequency, where the bandwidth $\delta \omega \sim \sigma^{-1}$. Consequently, for a pulse with a sub-threshold center frequency, photons in the higher energy end of the pulse spectrum could exceed the CDW gap and excite electron-hole pairs, a process that is further enhanced by nonlinear effects with a large laser fluence.

\begin{figure*}[]
\includegraphics[width=1.99\columnwidth]{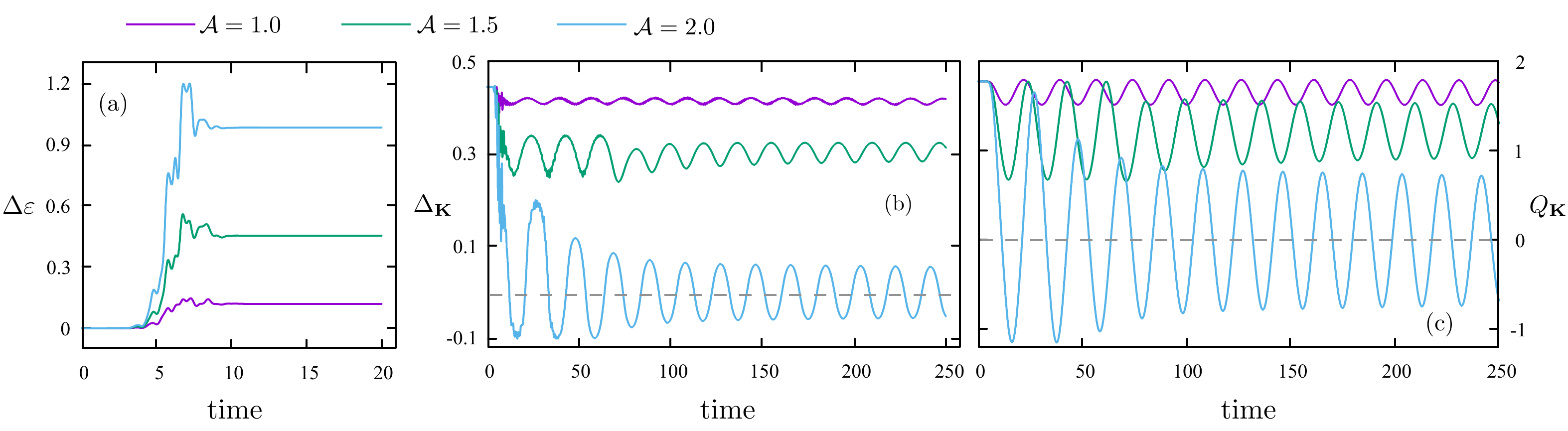}
\caption{\label{fig:n_amplitude} Time evoulution of (a) the injected energy per site~$\Delta \varepsilon(t) = [ E(t) - E_0 ]/N$, (b) the CDW order parameter $\Delta_{\mathbf K}(t)$, and (c) amplitude of staggered distortion $Q_{\mathbf K}(t)$, for laser pulse excitation with different amplitude $\mathcal{A}$. The width of the laser pulse is fixed at $\sigma = 2$,  the peak time is at $t_0 = 5$. The center frequency of the laser pulse is set at $\hbar\omega = 3$, which is well below the threshold $\hbar\omega_{\rm th} = 7.142$.}
\end{figure*}

A quantitative measure of the above-gap photoexcitation is the average energy (per site) $\Delta \varepsilon$ deposited to the system by the laser pulse. Fig.~\ref{fig:n_omega}(a) shows the injected energy $\Delta \varepsilon(t)$ as a function of time for various center frequencies. The pulse width is fixed at $\sigma = 2$, and the peak time is at $t_0 = 5$. Most of the energy injection occurs during the pulses width. The total energy remains nearly a constant after the pulse excitation, indicating a closed-system evolution under the coupled von~Neumann and Newton dynamics. Importantly, although the threshold frequency from the initial condition is $\hbar \omega_{\rm th} = 2 g \mathcal{Q} = 7.142$, significant energy transfer takes place already at $\hbar \omega \sim 3$. The overall energy deposition increases with the center frequency. 

To quantify the dynamics of the nonequilibrium CDW state, we introduce a time-dependent order parameter for the checkerboard density modulation 
\begin{eqnarray}
	\label{eq:CDW-delta1}
	\Delta_{\mathbf K}(t) = \frac{1}{N} \sum_i \langle \Psi(t) | \hat{n}_i | \Psi(t) \rangle e^{i \mathbf K \cdot \mathbf r_i}, 
\end{eqnarray}
where $\ket{\Psi(t)}$ is the single Slater determinant state, and the phase factor $\exp(i \mathbf K \cdot \mathbf r_i) = +1$ and $-1$ for sites in the A and B sublattice, respectively. Fig.~\ref{fig:n_omega}(b) shows the CDW order versus time for laser pulses of varying center frequencies.  The photoinduced nonequilibrium CDW states exhibit a diverse range of complex dynamical behaviors. Importantly, for most cases the laser pulse induces a prominent coherent oscillation of the CDW order which persists for a long time. This oscillation is accompanied by a nearly synchronized lattice dynamics as shown in Fig.~\ref{fig:n_omega}(c), where we plot the time dependence of the order parameter $Q_{\mathbf K}(t)$ that characterizes the overall staggered distortion
\begin{eqnarray}
	Q_{\mathbf K}(t) = \frac{1}{N} \sum_i Q_i(t) \, e^{i \mathbf K \cdot \mathbf r_i}.
\end{eqnarray}
The time evolution of the staggered lattice mode closely follows that of the CDW order parameter.  

As discussed above, sub-threshold excitation results in the reduction of CDW order already for frequencies as low as $\hbar\omega \sim 3$, which is well below the threshold $\hbar\omega_{\rm th} = 7.142$. Upon increasing the center frequency of laser pulses, more energy is injected into the CDW state, giving rise to a reduced CDW order on average. 
Moreover, the photoexcitation with a sub-threshold center frequency is further enhanced by an increasing laser intensity due to nonlinear effects. For example, FIG.~\ref{fig:n_amplitude} summarizes simulation results of laser excitations with a sub-threshold frequency $\hbar\omega = 3$ and varying fluences characterized by the parameter~$\mathcal{A}$. Upon increasing the laser fluence, more energy is deposited onto the system, which in turn results in a reduced average CDW order and an enhanced coherent oscillation. For laser excitations with a large enough fluence, e.g. $\mathcal{A} = 2.0$, even a pulse with sub-threshold frequency can completely melt the CDW order, as shown in the case of $\mathcal{A} = 2.0$ in FIG.~\ref{fig:n_amplitude}(b) and (c).

The mechanism of photoinduced suppression and melting of the checkerboard CDW order and the emergence of coherent oscillation can be understood as follows. Assuming negligible momentum of the incoming photons, the laser pulse excites a quasi-particle from the filled valence band $E^{-}_{\mathbf p}$ to the empty conduction band $E^+_{\mathbf p}$ of the initial CDW state, i.e. $|\pm\mathbf p\rangle \propto \hat{\gamma}^\dagger_{+, \mathbf p} \hat{\gamma}^{\,}_{-, \mathbf p} | {\rm CDW}_0 \rangle$, where $\hat{\gamma}_{\pm, \mathbf p}$ are the quasi-particle operators; see Appendix~\ref{app:bogoliubov} for details. The characteristic frequency of a pair of quasi-particle and quasi-hole is $ \hbar \nu_{\mathbf p} = E^+_{\mathbf p} - E^-_{-\mathbf p} = 2 \sqrt{ \varepsilon_{\mathbf p}^2 + (g \mathcal{Q})^2}$, where $\epsilon_{\mathbf p}$ is the dispersion relation of the square-lattice tight-binding model. In the absence of electron-electron scattering, the dynamics of a quasi particle-hole pair is an oscillatory motion with their natural frequency. Importantly, the photoexcited quasi-particles will modify the density matrix in Fourier space $\rho_{\mathbf p, \mathbf p + \mathbf K}$, which in turn contributes to the CDW order parameter 
\begin{eqnarray}
	\label{eq:delta_rho}
	& & \Delta_{\mathbf K}(t) = \sum_{\mathbf p} {\rho}_{\mathbf p, \mathbf p + \mathbf K}(t) \\
	& & \quad \qquad  = \sum_{\mathbf p, \mu\nu = \pm} C^{\mu\nu}_{\mathbf p} \langle \Psi(t) | \gamma^\dagger_{\mu, \mathbf p} \gamma^{\,}_{\nu, \mathbf p} |\Psi(t) \rangle. \nonumber
\end{eqnarray}
Here ${\rho}_{\mathbf p, \mathbf p + \mathbf K}(t) = \langle \Psi(t) | \hat{c}^\dagger_{\mathbf p + \mathbf K} \hat{c}^{\,}_{\mathbf p} | \Psi(t) \rangle$ describes the correlation of a electron-hole pair with a momentum difference $\mathbf K = (\pi, \pi)$ that characterizes the checkerboard charge modulation, and $C^{\pm,\pm}_{\mathbf p}$ are determined by coefficients that relate the quasiparticle operators to the electron operators; details can be found in Appendix~\ref{app:bogoliubov}.  The independent oscillations of different pairs with their respective natural frequencies give rise to a reduced CDW order due to destructive interferences in the summation of Eq.~(\ref{eq:delta_rho}).

Through the electron-phonon coupling, the oscillations of electron-hole pairs also initiate an oscillation of the checkerboard lattice distortion ${Q}_{\mathbf K}$ via the displacive excitation mechanism. Indeed, from the Fourier transform of Eq.~(\ref{eq:newton_eq}), the equation of motion for the staggered distortion is that of a simple Harmonic oscillator driven by an external force that is proportional to the CDW order parameter
\begin{eqnarray}
	\frac{d^2 Q_{\mathbf K}}{dt^2} + \Omega^2 Q_{\mathbf K} = \frac{g}{m} \Delta_{\mathbf K}
\end{eqnarray}
In equilibrium, a nonzero CDW order gives rise to a lattice distortion $Q_{\mathbf K} = (g/K) \Delta_{\mathbf K}$. As the CDW order is reduced from its initial value by the pulse excitation, the sudden shift to a new equilibrium results in the coherent phonon oscillation. 
The oscillation of the checkerboard lattice mode in turn drives the dynamics of electron-hole pairs ${\rho}_{\mathbf p, \mathbf p + \mathbf K}(t)$ which follows the dynamical equation
\begin{eqnarray}
	i\hbar \frac{d {\rho}_{\mathbf p, \mathbf p+\mathbf K}}{dt} = 2 \epsilon_{\mathbf p} {\rho}_{\mathbf p, \mathbf p+\mathbf K} 
	+ g Q_{\mathbf K} \left( {\rho}_{\mathbf p, \mathbf p} - {\rho}_{\mathbf p + \mathbf K, \mathbf p + \mathbf K} \right). \qquad
\end{eqnarray}
Although the oscillation of different electron-hole pairs favor their own natural frequencies, the dynamical coupling to a common checkerboard lattice oscillation promotes partial coherence among the various electron-hole pairs and locks them into a coherent oscillation of the CDW order that lasts for a long time.  

Depending on laser fluences, the oscillation amplitude can be seen to decay with time, although in most cases shown here the decay is rather slow. Since the system is isolated from any reservoir other than the initial short pulse excitation in our simulations, the damped oscillations of the CDW or staggered distortion are not caused by energy dissipations. Such dissipationless damping could result from a mechanism known as Landau damping where the energy of the collective mode, such as CDW order, is transferred to individual quasi-particle excitations. Indeed, Landau damping of coherent oscillations have been reported in quench dynamics of various symmetry-breaking phases including superconductivity~\cite{Volkov_1974,Barankov_2004,Yuzbashyan_2006a}, spin-density wave~\cite{Blinov_2017}, and CDW~\cite{yang24}. Detailed analysis of the coherent CDW oscillatory dynamics and their damping will be discussed elsewhere. 


\begin{figure}[]
\includegraphics[width=0.99\columnwidth]{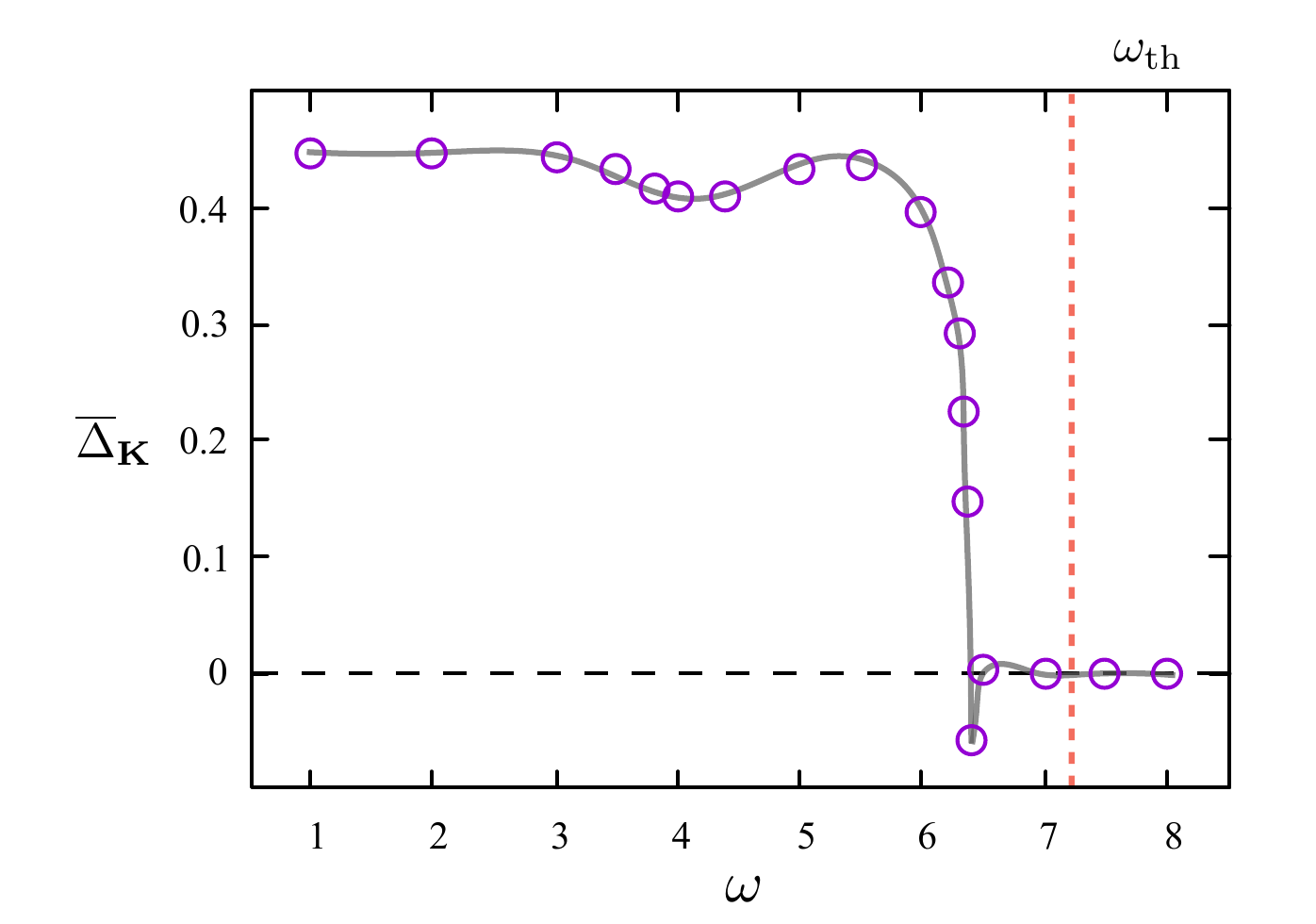}
\caption{\label{fig:omega_threshold} The late-stage CDW order parameter $\overline{\Delta}_{\mathbf K}$ averaged over several oscillation period versus the center frequency $\omega$ of the laser pulse. The dashed line indicates the threshold frequency $\omega_{\rm th} = 2 g \mathcal{Q} = 7.142$ determined from the energy gap of the initial CDW state.}
\end{figure}

The photo-induced melting of CDW order as a dynamical phase transition is summarized in FIG.~\ref{fig:omega_threshold} which shows the late-stage CDW order $\overline{\Delta}_{\mathbf K}$, averaged over several oscillation periods, as a function of the center frequency~$\omega$. 
Complete melting of CDW order occurs for frequencies $\hbar\omega \gtrsim 6.5$, which is below the threshold frequency $\hbar\omega_{\rm th} = 7.142$. Yet, the relation between the quasi-steady CDW order $\overline{\Delta}_{\mathbf K}$ and laser frequency is neither a sharp transition nor a smooth crossover. Immediately before the complete melting, the perturbed CDW state exhibits an intriguing charge-order inversion, indicated by a sharp dip at frequency $\hbar\omega \sim 6.4$ in the $\overline{\Delta}_{\mathbf k}$ versus $\omega$ curve shown in FIG.~\ref{fig:omega_threshold}. The corresponding time dependence of the CDW order $\Delta_{\mathbf K}(t)$ and the associated staggered distortion are shown in FIG.~\ref{fig:n_omega}(b) and (c); see the curves of $\hbar\omega = 6.4$. After the short pulse excitation, the CDW order quickly decays to zero, seemingly indicating a complete melting. However, instead of staying at zero, the CDW order flips sign and slowly grows to a steady state with a small oscillation around the average value. 

Similar photo-induced charge-order inversion phenomena were reported in previous theoretical studies of $Z_2$ type CDW/lattice order~\cite{Veenendaal_2016,Petrovic22,Ning20}. Experimentally, a laser-induced ultrafast reversal of combined excitonic order and lattice distortion has been observed in phonon coupled excitonic insulator Ta$_2$NiSe$_5$~\cite{Ning20,Guan23}. In general, the reversal occurs when the laser fluence is just large enough to induce a complete melting of CDW order (or the excitonic order). The physical picture of the charge-order reversal is as follows. The dephasing effect from different photoexcited electron-hole pairs quickly reduces the CDW order to zero when the pulse excitation is over. Through the electron-phonon coupling, the lattice distortion follows the vanishing CDW order and tends to zero itself. Yet, when the CDW order is being recovered partially, a nonzero lattice momentum carries the system across the zero and toward a state that is characterized by a CDW order of opposite sign.

Another interesting feature of the photo-induced dynamical transition in FIG.~\ref{fig:omega_threshold} is a broad dip at the mid-gap frequency $\omega_r \sim \omega_{\rm th}/2$. The reduced CDW order at this sub-threshold frequency indicates an enhanced photo-excitation. A similar intensified photo-induced dynamics was also reported in CDW states of the 1D $t$-$V$ model~\cite{Seo18} and BCS superconductors~\cite{Matsunaga14,Krull2016}. Since there is no in-gap state in the initial homogenous CDW insulator, this mid-gap dip cannot be ascribed to a linear resonant absorption. On the other hand, as the resonant frequency is roughly half the CDW band gap, photo-excitations of quasi-particles can be achieved through a nonlinear two-photon process with $2 \hbar \omega_r \sim \epsilon_{\rm gap}$. Moreover, since the electron DOS exhibits a divergence at the band edge $\varrho(E) \sim E / \sqrt{E^2 - (\epsilon_{\rm gap} / 2)^2}$, the resonance at the mid-gap $\omega_r$ results from the heightened two-photon absorption assisted by an enhanced electron density of states at the edge of the CDW band gap.

\begin{figure*}[t]
\includegraphics[width=1.99\columnwidth]{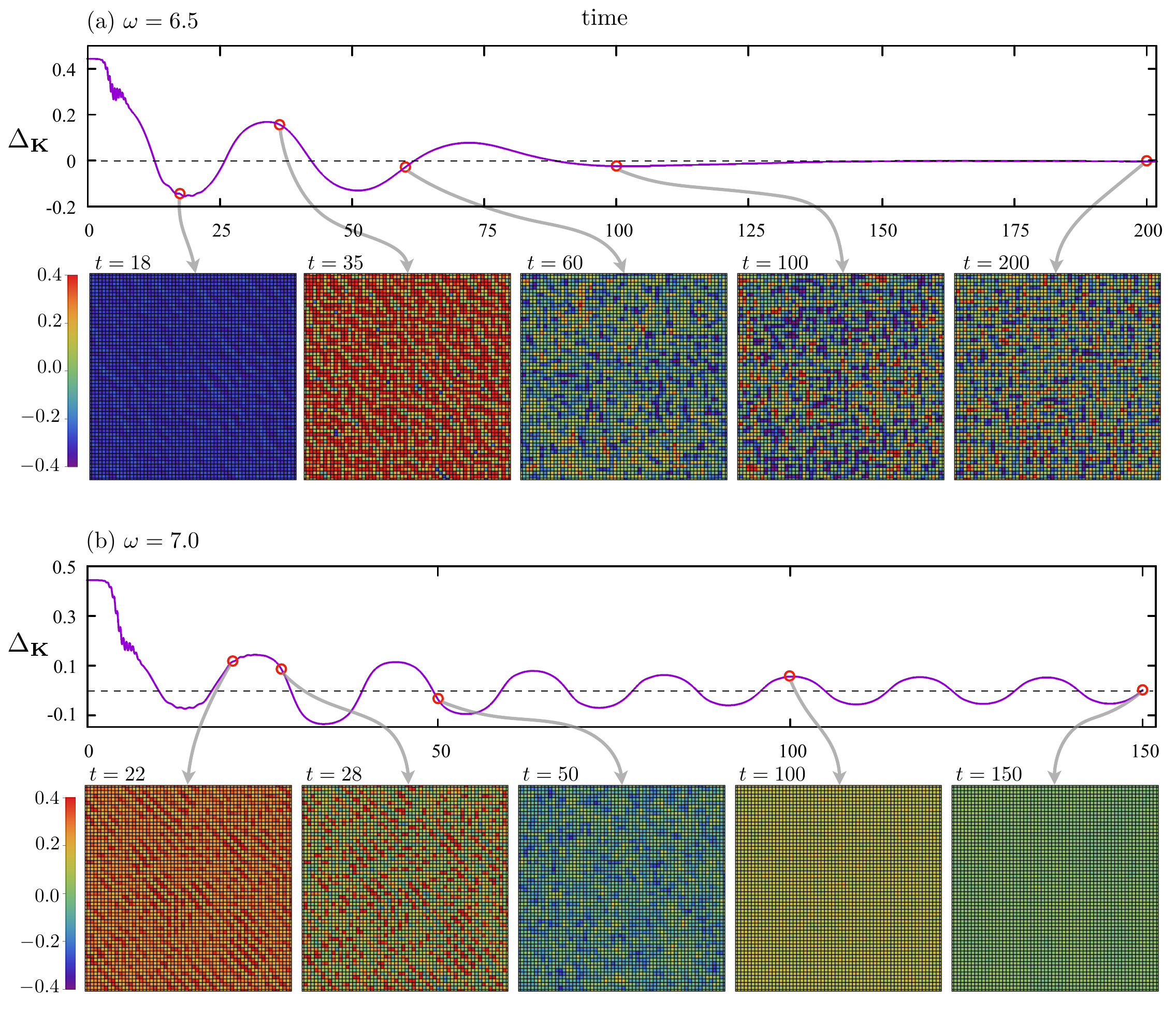}
\caption{\label{fig:snapshots1}
The time dependence of global CDW order $\Delta_{\mathbf K}(t)$ and representative CDW configurations during the photo-excitation process for laser pulses with a frequency of (a) $\hbar \omega = 6.5$ and (b) $\hbar \omega = 7.0$. }
\end{figure*}

\section{Dynamical Inhomogeneity and Pattern Formation}

\label{sec:pattern}

For the cases immediate preceding the complete melting, e.g. $\omega \sim 6.4$ or 6.5, the coherent oscillation amplitudes exhibit a significant damping. Yet, for laser frequency above the threshold such as $\omega = 7.0$, the complete melting of the CDW order is followed by a pronounced coherent oscillation that lasts for a long time. Although damped oscillations can be understood as arising from the Landau damping mechanism discussed above, the strong damping of the CDW order at excitation frequencies near the threshold is related to the emergence of spatial inhomogeneity.

The real-space simulations discussed in Sec.~\ref{sec:model} can provide information about the spatial inhomogeneity. In particular, to characterize the emergence of nonuniform CDW states, we define the following local CDW order parameter
\begin{eqnarray}
\label{eq:local_CDW}
    \phi(\mathbf r_i) = \Bigl(n_i - \frac{1}{4}\sum_j\phantom{}^{'} n_j \Bigr) \exp\left({i \mathbf K \cdot \mathbf r_i}\right). 
\end{eqnarray}
Here, the prime indicates that the summation is restricted to the four nearest neighbors of site-$i$. This quantity measures the difference in electron density between a given site and its nearest neighbors. The phase factor, $\exp\left({i \mathbf K \cdot \mathbf r_i}\right)=\pm 1$, is introduced to account for the ultra-short range checkerboard modulation within a CDW domain. A homogeneous CDW state is thus described by a constant local order parameter, and any inhomogeneity is manifested as a spatially varying $\phi(\mathbf r)$ field.

\begin{figure*}[t]
\includegraphics[width=1.99\columnwidth]{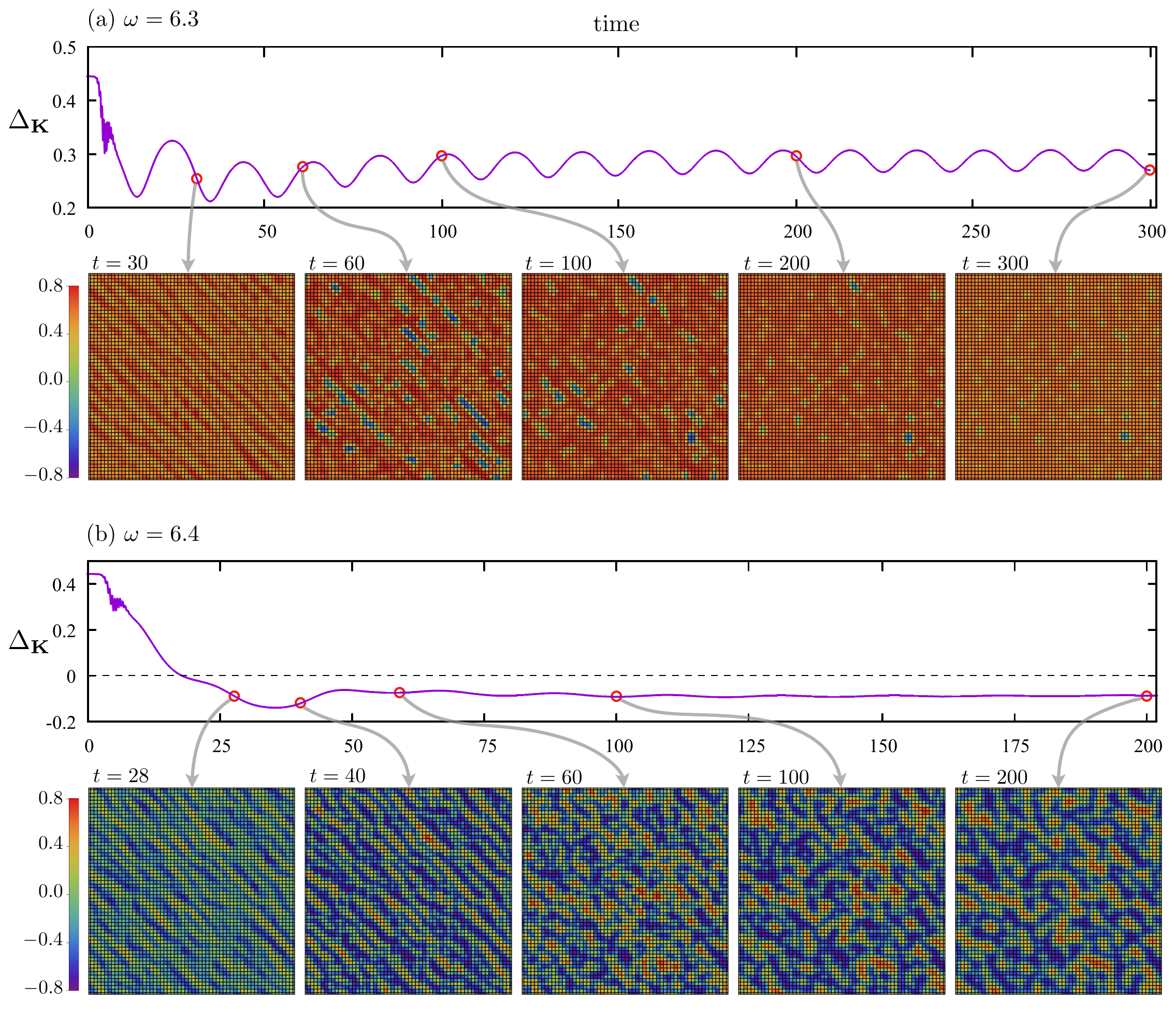}
\caption{\label{fig:snapshots2}
The time dependence of global CDW order $\Delta_{\mathbf K}(t)$ and representative CDW configurations during the photo-excitation process for laser pulses with a frequency of (a) $\hbar \omega = 6.3$ and (b) $\hbar \omega = 6.4$. }
\end{figure*}

First we compare the two cases where laser excitations result in complete melting of the CDW order. FIG.~\ref{fig:snapshots1} shows the time dependence of the global CDW order parameter $\Delta_{\mathbf K}(t)$ and the real-space configuration of the local CDW order $\phi(\mathbf r, t)$ at various times for the two frequencies $\hbar\omega = 6.5$ and~7.0. In the former case, the global CDW order exhibits a damped oscillation with the oscillation amplitude decays to zero at late times; see FIG.~\ref{fig:snapshots1}(a). As the snapshot at $t = 18$ shows, the CDW order parameter $\phi(\mathbf r)$ remains roughly homogeneous when $\Delta_{\mathbf K}$ changes to the negative sign. Yet, when the global CDW order bounces back to the positive side at $t = 35$, significant inhomogeneity has developed in the CDW state. As the damped oscillation reaches a steady state with a nearly vanishing global CDW order $\Delta_{\mathbf K} \approx 0$ at times $t \gtrsim 125$, the system stays in a highly inhomogeneous state with a standard deviation of the local CDW order parameter as high as $\sigma_{\phi} = \sqrt{\langle \phi_i^2 \rangle - \langle \phi_i \rangle^2} \sim 0.3$. It is worth noting that, since the snapshot shows the spatial configuration of the local order-parameter field $\phi(\mathbf r)$ (instead of the charge density itself), the observed inhomogeneity corresponds to a super-modulation of charge density on top of the underlying $(\pi, \pi)$ checkerboard  charge pattern.

For laser excitation with a center frequency $\hbar \omega = 7.0$, the snapshots at times $t = 22$ and 28 also exhibit noticeable inhomogeneity; see FIG.~\ref{fig:snapshots1}(b). In particular, the inhomogeneous CDW states at $t = 22$ clearly shows a pattern of stripes running along the  $y = -x$ diagonal direction, which is perpendicular to the direction of electric field of the laser pulse. The initial CDW order is melted in the sense that the time-averaged global CDW order tends to zero at late times. However, a pronounced coherent oscillation of both the CDW order and staggered lattice distortion remains. Moreover, in stark contrast to the previous melting scenario which ends with a rather disordered CDW state,  the local CDW order parameter field $\phi(\mathbf r_i)$ in this dynamical state is found to be rather homogeneous, as demonstrated in the snapshots of $t = 100$ and 150 in FIG.~\ref{fig:snapshots1}(b). 

The emergence of this dynamical regime in the $\hbar \omega = 7.0$ case can be understood as follows. The pump pulse with a larger center frequency produces electron-hole pairs at higher energies. While such excitations quickly leads to the destruction of the static CDW order, a coherent phonon oscillation $Q_{\mathbf K}$ with a larger amplitude is also generated. The incoherent dynamics of photo-excited quasi-particles manifests itself not only in the time domain, as described by the Landau-damping mechanism, but also in the spatial dimension which leads to the inhomogeneous CDW states in the early stage of the melting process. Yet, the enhanced oscillations of the phonons reinforce the coherence of the electron-hole pairs both in temporal and spatial domains, giving rise to a sustained oscillation of the CDW order in a relatively homogeneous state. Indeed, although there is no static global CDW order after averaging over time, the relatively homogeneous CDW state shown in FIG.~\ref{fig:snapshots1}(b), sustained by a coherent phonon oscillation, can be viewed as a dynamical counterpart of the conventional CDW state.

As discussed in Sec.~\ref{sec:melting_order}, the photo-excited electron-hole pairs have a relative momentum $\mathbf K = (\pi, \pi)$, which would modify the initial checkerboard CDW pattern. The incoherence of the electron-hole pairs only reduces the amplitude of checkerboard CDW, which is expected to remain spatially homogeneous. A spatial modulation of the particle density with a wave vector $\mathbf q$ requires electron-hole correlations with the same relative momentum, i.e. 
\begin{eqnarray}	
	\label{eq:Delta_p}
	\Delta_{\mathbf q}(t) = \sum_i \rho_{ii}(t)\, e^{i \mathbf q \cdot \mathbf r_i} = \sum_{\mathbf p} \rho_{\mathbf p, \mathbf p + \mathbf q}(t). 
\end{eqnarray}
Consequently, the emergence of spatial inhomogeneity indicates the spontaneous generation of electron-hole pairs $\rho_{\mathbf p, \mathbf p + \mathbf q} = \langle c^\dagger_{\mathbf p + \mathbf q} c^{\,}_{\mathbf p} \rangle$ with a relative momentum $\mathbf q \neq \mathbf K$, i.e. different from the initial checkerboard $\mathbf K$. 
The generation of such additional density modulations is often achieved through an instability mechanism of pattern formation. More precisely, the spatial patterns arise from the amplification of initial infinitesimal density fluctuations of certain wave vectors through nonlinear effects of the dynamical evolution.  

This scenario is illustrated in the two cases shown in FIG.~\ref{fig:snapshots2}. For the first case, the initial CDW order is reduced by the laser pulse with $\hbar \omega = 6.3$, yet a finite static time-averaged global CDW order parameter $\overline{\Delta}_{\mathbf K} \sim 0.28$ remains at late times. Importantly, a spatially inhomogeneous CDW state with stripes running along the $y = -x$ diagonal can be seen at $t = 30$. At time $t = 60$ in FIG.~\ref{fig:snapshots2}(a), there are several diagonal streaks where the local CDW order parameter $\phi_i$ changes sign. Interestingly, such stripe pattern is similar to those observed in FIG.~\ref{fig:snapshots1} with larger laser frequencies. The additional density modulations $\Delta_{\mathbf q}$ are characterized by wave vectors $\mathbf q \parallel \mathbf E$, parallel to the electric field direction.  As the system settles into a self-sustained coherent oscillation with the phonons, the diagonal patterns are gradually suppressed. This is another example of restoration of partial homogeneity by the coherent dynamics. However, a small residual inhomogeneity remains even at late times. 

The nonequilibrium CDW state induced by a laser pulse with $\hbar\omega = 6.4$ not only shows the phenomenon of charge-order reversal, but also exhibits the a pronounced pattern formation as shown in FIG.~\ref{fig:snapshots2}(b). Contrary to the previous $\hbar\omega = 6.3$ case, the global CDW order $\Delta_{\mathbf K}(t)$ shows nearly indiscernible oscillations and quickly settles to a steady-state value (with a sign opposite to the initial CDW state). This lack of late-time coherent oscillation is also intimately related to the emergent dynamical inhomogeneity. Indeed, stripe modulations of the local CDW order can be seen at times as early as $t = 28$ and 40 in the charge-inverted state. As the system further evolves, more complicated patterns emerge with an even stronger modulation of the $\phi$-field. 

\begin{figure}[]
\includegraphics[width=86mm]{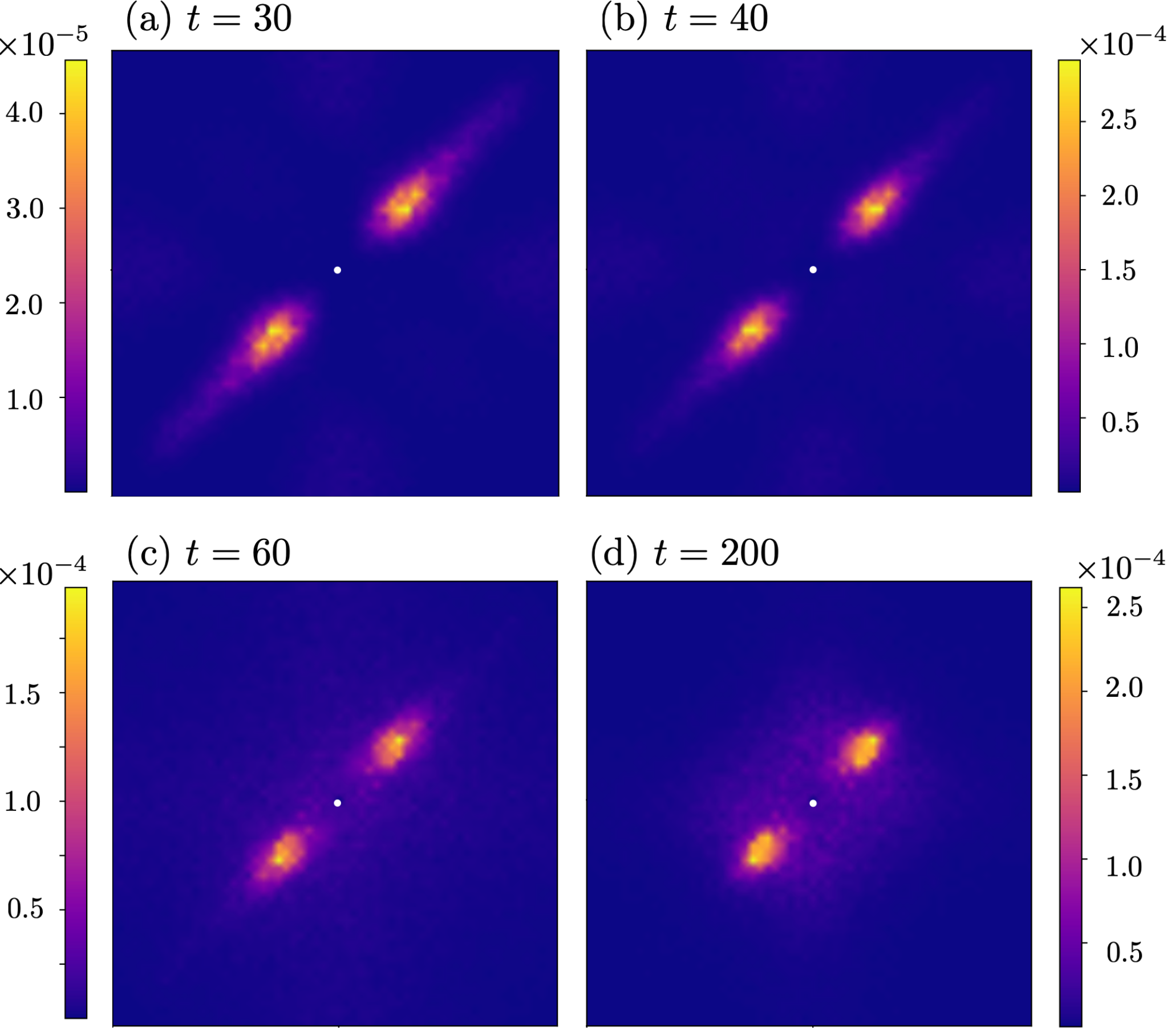}
\caption{\label{fig:struc_fac} Structure factors $S(\mathbf q, t)$ at various time steps after the pump-probe excitations with a laser frequency $\hbar\omega=6.4$. The peak time is $t_0 = 5$ and pulse width is $\sigma = 2$.  The simulated system size is $N = 60 \times 60$. The results are obtained by averaging over 20 independent von~Neumann dynamics simulations. The white dot at $\mathbf K = (\pi, \pi)$ corresponds to a dominant checkerboard CDW order. The scale of the color bars in all panels are chosen to highlight the emergent unstable modes.}
\end{figure}

It is worth noting that, in terms of the on-site electron density $\rho_{ii}$, these stripe patterns correspond to a super modulation of charge density on top of an underlying ultrashort range checkerboard charge pattern. The more complicated patterns at late times, e.g. $t = 200$, seem to originate from the breaking up of the original stripes. To further characterize this inhomogeneous state, we compute the structure factor of the charge density
\begin{eqnarray}
	S(\mathbf q, t) =  \left| \Delta_{\mathbf q}(t) \right|^2,
\end{eqnarray}
where $\Delta_{\mathbf p}(t)$ is the density modulation defined in Eq.~(\ref{eq:Delta_p}). FIG.~\ref{fig:struc_fac} shows the structure factors of the nonequillibrium CDW states induced by a $\hbar \omega = 6.4$ laser pulse at various times in a pump-probe setup. The peak time of the pump pulse is at $t_0 = 5$, and the pulse width is $\sigma = 2$. The sharp peak at $\mathbf K = (\pi, \pi)$ that can be seen at structure factors of different times indicates that a finite checkerboard CDW persists throughout the post-pump evolution. The photo-induced pattern formation, in the form of super density modulation, manifests itself in the emergence of two broad diffusive peaks along the diagonal direction at $\mathbf q^*_{\pm} = \mathbf K \pm (2\pi / \lambda) \hat{\mathbf e}$, where $\hat{\mathbf e} = (\hat{\mathbf x} + \hat{\mathbf y})/\sqrt{2}$ is also the electric field direction, and $\lambda$ can be viewed as a characteristic, or average, period of the stripe modulations. 

The two diffusive peaks are also highly anisotropic in shape; the spreading is predominately along the same diagonal direction $\hat{\mathbf e}$. The widths of the diffusive peaks parallel and perpendicular to the diagonal direction provide a measure of corresponding correlation lengths $\xi_{\parallel}$ and $\xi_{\perp}$. The breaking up of the stripes at late times, e.g. $t = 100$ and 200 in FIG.~\ref{fig:snapshots2}(b), indicates a decreased correlation in the perpendicular direction, which is consistent with the fact that $\xi_{\perp}$ seems to remain roughly the same. On the other hand, the longitudinal correlation seems to be enhanced, as $\xi_{\parallel}$ inferred from the structure factor is increased at late times.

A possible scenario for the emergence of the stripe patterns is the decay of the checkerboard CDW order parameter into a pair of such unstable modes at $\mathbf q^*_{\pm}$ through the parametric instability mechanism. The standard scenario would be that the checkerboard coherent phonon oscillation $Q_{\mathbf K}(t)$, acting as pump wave, couples to the two unstable $Q_{\mathbf q^*_\pm}$ modes through an effective three-wave nonlinear coupling mediated by the electrons. 
However, for general modulation period $\lambda$, the momentum is not conserved in this process, $\mathbf q^*_+ + \mathbf q^*_- \neq \mathbf K$. This consideration thus rules out the standard parametric instability as the mechanism of the stripe patterns observed in our simulations. On the other hand, the fact that the unstable modes appear in pairs still strongly indicates a parametric-like mechanism, possibly through an intermediate pump wave of electronic origin.  A detailed study of the instability mechanism will be left for future.

\section{Conclusion and Outlook}

\label{sec:conclusion}

In summary, we have conducted a comprehensive investigation into the ultrafast dynamics of CDW states in a semi-classical Holstein model in a pump-probe setup.  For the square-lattice model at half-filling, the ground state of the Holstein model exhibits a checkerboard CDW order characterized by the wave vector $\mathbf K = (\pi, \pi)$. Within the semiclassical approximation where phonons are treated as classical degrees of freedom, an efficient real-space method, based on von~Neumann equation for electrons and Newton equation for lattice dynamics, is developed to simulate the pump-probe process. Our extensive simulations uncover intriguing dynamical behaviors such as the melting of CDW order with or without a late-time coherent oscillations, mid-gap nonlinear resonance, and reversal of CDW order near the melting threshold. But most importantly, our large-scale real-space approach reveals the crucial role of dynamical inhomogeneity and pattern formation in the ultrafast dynamics of nonequilibrium CDW states. 

\begin{figure}
\includegraphics[width=80mm]{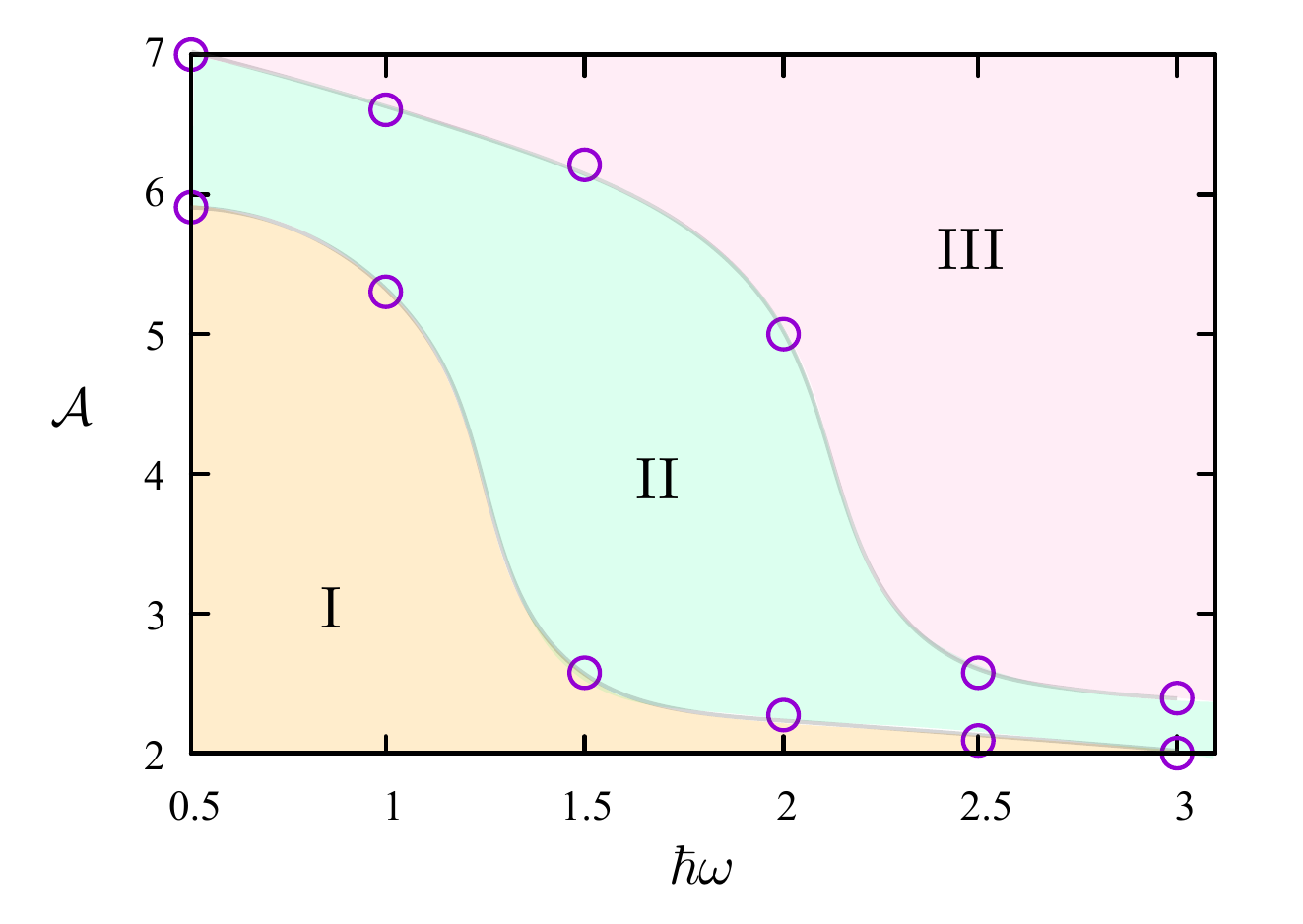}
\caption{\label{fig:phase_diagram} Phase diagram showing three dynamical regimes: (I)~Coherent oscillations of homogeneous CDW and staggered lattice distortion, with a finite time-averaged order parameter. (II)~The emergence of pattern formation of the local CDW order parameter. (III)~Dynamical CDW order sustained by a checkerboard coherent phonon oscillations with a zero time-averaged order parameter $\overline{\Delta}_{\mathbf K} = 0$.
}
\end{figure}

Our main results are summarized in the phase diagram shown in FIG.~\ref{fig:phase_diagram}. Depending on the laser fluence $\mathcal{A}$ and center frequency $\omega$ (which determines the photon energy), there are three dynamical regimes of the photo-induced nonequilibrium CDW states in the phase diagram. They can be characterized by two parameters: the time-averaged global CDW order $\overline{\Delta}_{\mathbf K}$ at late times and the standard deviation $\sigma_{\phi}$ of the local CDW order parameter $\phi_i$. In dynamical phase~I, induced by pump pulses of a small photon energy and a moderate fluence, the phtoexcitation suppresses the CDW order and initiates a coherent oscillation. Yet, the averaged CDW order remains finite, $\overline{\Delta}_{\mathbf K} \neq 0$, with a negligible inhomogeneity $\sigma_\phi \approx 0$. Upon increasing the laser frequency, dynamical phase~II shows pronounced spatial inhomogeneity and the emergence of pattern formation; the CDW states in this regime are mostly characterized by $\sigma_{\phi} \neq 0$. Finally, for the third dynamical regime at large fluence and frequency, the static CDW order is melted down in the sense that the time-averaged order vanishes $\overline{\Delta}_{\mathbf K} \approx 0$. Moreover, a strong coherent phonon oscillation partially restores the homogeneity. The nonequilibrium states in this regime are described by a dynamical global CDW order parameter $\Delta_{\mathbf K}(t) \sim \cos(\Omega t + \varphi_0)$. Since there is {\em no} static CDW order after averaging over oscillation periods, phase~III can be viewed as a dynamical version of the conventional CDW phase. 

Previous works on similar CDW order in 1D systems already indicated the importance of inhomogeneity during the melting process, for example, the soliton-driven melting scenario~\cite{Hashimoto17,Seo18}. A similar result of photo-induced dynamical inhomogeneity was reported in the case of 2D checkerboard CDW, although the electron degrees of freedom are modeled by a classical two-level molecular bond charges~\cite{Veenendaal_2016}. The inhomogeneity, especially due to the proliferation of topological defects, is a well-studied theme already in equilibrium physics and  thermal-induced out-of-equilibrium processes such as the conventional Kibble-Zurek mechanism. It is worth noting that the emergence of inhomogeneity in such scenarios is due to the lack of spatial coherence during a relaxation process.  Yet, our work here highlights another route to inhomogeneity through a pattern formation instability, a defining feature of a nonequilibrium system with a highly nonlinear dynamics.

The checkerboard CDW which breaks the $Z_2$ sublattice symmetry is perhaps one of the simplest symmetry-breaking phases. Our work shows that pattern formation instability could take place even in a many-body quantum system with such relatively simple Ising order. While the CDW order is stabilized by electron-phonon coupling in the Holstein model considered in this work, similar pattern formation phenomena were also observed in the quench dynamics of checkerboard CDW of pure electronic origin~\cite{yang24}. Previous theoretical works have also reported pattern-formation instability as the source of Cooper pair turbulence in BCS type superconductors~\cite{Dzero_2009,Chern19}. It is expected that pattern formation would be rather ubiquitous in ultrafast dynamics of quantum systems with more complex order parameters. Although the mechanisms and dynamics of pattern formation have been extensively studied in many branches of classical physics, a comprehensive picture of similar phenomena in quantum many-body systems is still lacking. It is likely that a complete theoretical description would require techniques and ideas beyond the classical theories. For example, preliminary analysis already points out the importance of nonlinear couplings between electron-hole pairs and collective behaviors (e.g. order-parameter dynamics) in the instability mechanism. Further investigations, both theoretical and experimental, will be needed to explore this exciting new field.

\appendix

\section{Quasiparticles and energy gap of a charge density wave state}

\label{app:bogoliubov}

We consider a staggered lattice distortion described by
\begin{eqnarray}
	Q_i = \mathcal{Q} \exp(i \mathbf K \cdot \mathbf r_i),
\end{eqnarray}
where $\mathcal{Q}$ is the amplitude of lattice distortion and $\mathbf K = (\pi, \pi)$ is the wave vector of the checkerboard pattern. By introducing electron creation/annihilation operators in momentum space, e.g. $\hat{c}^\dagger_{\mathbf p} = \frac{1}{\sqrt{N}} \sum_i \hat{c}^\dagger_i \,e^{i \mathbf p \cdot \mathbf r_i}$, the electron Hamiltonian $\hat{\mathcal{H}}_{\rm e} + \hat{\mathcal{H}}_{\rm eL}$ can be expressed as
\begin{eqnarray}
	\hat{\mathcal{H}}_{\rm CDW} = \sum_{\mathbf p} \hat{\bm c}^\dagger_{\mathbf p} H(\mathbf p) \hat{\bm c}^{\,}_{\mathbf p}, 
\end{eqnarray}
where the summation is restricted to the reduced Brillouin zone, $\hat{\bm c}^{\,}_{\mathbf p} = (\hat{c}^{\,}_{\mathbf p}, \, \hat{c}^{\,}_{\mathbf p + \mathbf K})^{t}$ is a column vector of the electron operators, and $H(\mathbf k)$ is the one-particle Hamiltonian given by
\begin{eqnarray}
	H(\mathbf p) = \left( \begin{array}{cc} \epsilon_{\mathbf p} & - g \mathcal{Q} \\ -g \mathcal{Q} & -\epsilon_{\mathbf p} \end{array} \right).
\end{eqnarray}
The diagonal elements are given by the dispersion relation of the square-lattice tight-binding model
\begin{eqnarray}
	\epsilon_{\mathbf p} = -2t_{\rm nn}(\cos q_x + \cos q_y),
\end{eqnarray}
and we have used the relation $\epsilon_{\mathbf p + \mathbf K} = -\epsilon_{\mathbf p}$ in the matrix equation for $H$. The CDW Hamiltonian can be straightforwardly diagonalized by the Bogoliubov transformation. To this end, we introduce quasi-particle operators
\begin{eqnarray}
 \label{eq:bogoliubov}
     \hat{\gamma}^{\dag}_{+,\mathbf{q}} = u_{\mathbf{q}}\hat{c}^{\dag}_{\mathbf{q}}-v_{\mathbf{q}}\hat{c}^{\dag}_{\mathbf{q}+\mathbf{K}}, \quad
    \hat{\gamma}^{\dag}_{-,\mathbf{q}} = v_{\mathbf{q}}\hat{c}^{\dag}_{\mathbf{q}} + u_{\mathbf{q}}\hat{c}^{\dag}_{\mathbf{q}+\mathbf{K}}, \qquad
\end{eqnarray}
where the transformation coefficients are given by
\begin{eqnarray}
	& & u_{\mathbf p} = \frac{1}{\sqrt{2}} \left(1 + \frac{\epsilon_{\mathbf p}}{\sqrt{\epsilon_{\mathbf p}^2 + (g \mathcal{Q})^2}} \right)^{1/2}, \nonumber  \\
	& & v_{\mathbf p} = \frac{1}{\sqrt{2}} \left(1 - \frac{\epsilon_{\mathbf p}}{\sqrt{\epsilon_{\mathbf p}^2 + (g \mathcal{Q})^2}} \right)^{1/2}.
\end{eqnarray}
The diagonalized Hamiltonian becomes
\begin{eqnarray}
	\hat{\mathcal{H}}_{\rm CDW} = \sum_{\mathbf p}\sum_{\mu = \pm} E^\mu_{\mathbf p} \hat{\gamma}^\dagger_{\mu, \mathbf p} \hat{\gamma}^{\,}_{\mu, \mathbf p}.
\end{eqnarray}
where the quasi-particle energies are
\begin{eqnarray}
	E^{\pm}_{\mathbf p} = \pm \sqrt{ \epsilon_{\mathbf p}^2 + (g \mathcal{Q})^2}
\end{eqnarray}
A spectral gap $\epsilon_{\rm gap} = 2 g \mathcal{Q}$ is opened in the excitation spectrum.  At half-filling, the electron energy is obtained by filling up all negative energy states, giving rise to an energy density of the CDW state
\begin{eqnarray}
	\varepsilon(\mathcal{Q}) =- \frac{1}{N} \sum_{\mathbf p} \sqrt{ \epsilon_{\mathbf p}^2 + (g \mathcal{Q})^2} + \frac{1}{2} K \mathcal{Q}^2,
\end{eqnarray}
The order parameter of the staggered lattice distortion~$\mathcal{Q}$ is determined from the minimum energy condition $\partial \varepsilon / \partial \mathcal{Q} = 0$.  

The electron-hole correlation function $\rho_{\mathbf p, \mathbf p+\mathbf K} = \langle \hat{c}^\dagger_{\mathbf p + \mathbf K} \hat{c}^{\,}_{\mathbf p} \rangle$ can be expressed in terms of quasi-particle operators, as shown in Eq.~(\ref{eq:delta_rho}), with the following expansion coefficients
\begin{eqnarray}
 \label{eq:order_parameter_quasi}
 \begin{aligned}
    &C^{+,+}_{\mathbf{p}} = -u_{\mathbf{p}}v_{\mathbf{q}},\quad C^{-,+}_{\mathbf{p}} = u^2_{\mathbf{q}},\\
    &C^{-,-}_{\mathbf{p}} = u_{\mathbf{p}}v_{\mathbf{q}},\quad C^{+,-}_{\mathbf{p}} = -v^2_{\mathbf{q}},
\end{aligned}
\end{eqnarray}

\bigskip

\begin{acknowledgments}
The authors thank Yang Yang for useful discussions. The work was supported by the US Department of Energy Basic Energy Sciences under Contract No. DE-SC0020330. L. Yang acknowledges the support of Jefferson Fellowship by the Jefferson Scholars Foundation. The authors also acknowledge the support of Research Computing at the University of Virginia.
\end{acknowledgments}

\bibliography{ref}
\end{document}